# STUDENTS' PATTERNS OF INTERACTION WITH A MATHEMATICS INTELLIGENT TUTOR: LEARNING ANALYTICS APPLICATION

Anita Dani

Faculty of Mathematics, Higher Colleges of Technology, United Arab Emirates

## ABSTRACT

*Purpose: The purpose of this paper is to determine potential identifiers of students' academic success in foundation mathematics course by analyzing the data logs of an intelligent tutor.*

*Design/ methodology/approach: A cross-sectional study design was used. A sample of 58 records was extracted from the data-logs of the intelligent tutor, ALEKS. This data was triangulated with the data collected from surveys. Two-step clustering, correlation and regression analysis, Chi-square analysis and paired sample t-tests were applied to address the research questions.*

*Findings: The data-logs of ALEKS include information about number of topics practiced and number of topics mastered by each student. Prior knowledge and derived attribute, which is the ratio of number of topics mastered to number of topics practiced(denoted by the variable m top in this paper) are found to be predictors of final marks in the foundation mathematics course with $R^2 = 42\%$.*

*Students were asked to report their preferred way of selecting topics as either sequential or random. Results of paired sample t-test demonstrated that the students who selected topics in a sequential manner were able to retain their mastery of learning after the summative assessment whereas the students who chose topics randomly were not able to retain their mastery of learning.*

*Originality and value: This research has established three indicators of academic success in the course of foundation mathematics which is delivered using the intelligent tutor ALEKS. Instructors can monitor students' progress and detect students at-risk who are not able to attain desired pace of learning and guide them to choose the correct sequence of topics.*

## 1. INTRODUCTION

Intelligent tutoring systems can provide individual tutoring, instant feedback on learning and flexibility to learn at own pace. Intelligent tutors provide interactive and personalized learning environment for students which can facilitate student centered learning. Since these software tools are now available on mobile devices, such as iPads, their adaption in teaching and learning is changing the teacher centered learning into student centered learning. Though this change is promising it also implies that students are expected to develop self-regulatory abilities [30,31]. A key feature of these software systems is their ability to record and store details of each learning activity that happens when a student interacts with the system. Such digital data can be analyzed and interpreted which can reveal some useful information regarding students' study habits and their progress [22]. Learning analytics is a research discipline which provides framework for analysis of these system-generated large data logs in order to understand learning activities [17,39,41]. Use of such systems and reports generated by methods of learning analytics empower faculty to engage students in authentic learning opportunities and can increase student participation and motivation [10].

ALEKS is a web-based intelligent tutoring system, which is used in one of the higher education institutes in the United Arab Emirates. The name ALEKS is an acronym for Assessment and Learning in Knowledge Spaces, which indicates that this software provides learning opportunities through frequent formative assessments. ALEKS encompasses theories of learning [2], assessment and intelligent tutors. It provides opportunity to set learning goals, sufficient practice material on each conceptual unit to master the concept and administers frequent formative assessments to provide feedback on learning. It mimics the ability of an expert teacher and determines correctness of a student's next response based on his or her current response. Due to this ability it can confirm whether the student has mastered a conceptual unit and whether the student has retained the achieved mastery.

Unlike traditional paper-based mathematics textbooks, these systems provide flexibility of choosing topics from any chapter. Two topics from different chapters may be linked on the basis of theory of knowledge spaces [14], but they need not be solved using the same problem solving strategy. Many students may attempt to master many topics from the course in a short time, without organizing their learning tasks and may choose topics randomly. If a student is not able to switch from one problem solving strategy to another one, she may fail to master a new topic or may fail to retain mastery of her learning. On the other hand, some students choose their topics sequentially where they master all units from one chapter and then move to the next chapter.
Aim of our research is to determine if sequencing of topics during learning has any impact on the ability to retain mastery of learning. The second aim is to determine other predictors of students' academic success based on her patterns of interactions with the system. This examination is based on the student's learning patterns obtained from the data logs generated by ALEKS as well as data collected from short surveys.

The paper is organized as follows: A brief review of literature is given in the next section, which is followed by the description of intelligent tutor and the research context. Research methodology is presented in the section 5. Section 6 contains results and discussion. The paper concludes with recommendations and direction for future research.

## 2. LITERATURE REVIEW

This section contains a brief review of literature on instruction theories, use of intelligent tutors in teaching mathematics and learning analytics.

### 2.1. Theory of instructions in mathematics

Importance of sequencing learning tasks during instruction and during revision has been researched and emphasized by many researchers [21,,33,34,35,36,43]. The instructional strategy in which the learning tasks are presented in a group and delivered over a longer period is known as *blocked or massed* practice [25]. In this type of instructions students are expected to solve questions which require the same problem solving strategy. Students may master a concept quickly during blocked practice but may fail to retain the mastery as well as may not be able to develop meta-cognitive strategy. As discussed in [34], with blocked practice, students know the problem solving strategy beforehand. Many a times, students do not read the complete problem statement and solve the questions mechanically. When they are assessed in a formative assessment, they may not be able to identify the correct problem solving strategy. *Random practice* refers to an instructional strategy in which the tasks given for practice require all different problem solving strategies. Students may not be able to connect links between any two successive tasks and as a result, may not be able to learn during such practice sessions. Many studies have proposed a strategy that balances between these two methods, which is termed as *interleaved practice.* During this practice session, students practice one type of task for a short duration then switch to a different task for another short duration. This practice develops cognitive and meta-cognitive abilities in students and helps them to retain their mastery[25,34,35]. This strategy is more effective than the blocked or random strategy, since students are exposed to different types of problems and introduced appropriate strategies for each type. During the practice session they are expected to map the correct strategy for each problem.

Assessments are a source of learning therefore assessment strategies and instruments should have cognitive as well as motivational purpose [26,28,37,38,40]. Computer based assessments have several advantages over paper based assessments [28]. They are available online providing access to any number of students anytime and anywhere. They provide a wider range of assessment techniques than the paper based assessments, such as inclusion of graphics and multimedia. Student's responses to the assessment questions can be numbers and texts as well as hotspot clicking. Evaluation and feedback is given instantly by these systems [7]. More importantly, software can generate questions randomly from a large question bank. Randomly generated tasks in the assessment can be useful to assess students' knowledge and ability to apply the correct problem solving strategy [35]. This type of web-based assessment software can foster the student-centered learning by engaging students in meaningful learning activities and by fostering skills of independent learning [13].

Web-based assessment and practice lead students to have more control over their work and their effort as they get the immediate feedback and instant scoring [12,27,29,30,32]. Though effectiveness of intelligent tutors to teach algebra has been confirmed by many researchers [8,12,18]. Some researchers also found that non-cognitive factors, such as affect towards intelligent tutors have an impact on students' learning [19]. Though timely feedback is a powerful tool which can guide students in the correct direction, its effect is moderated by prior knowledge [16]. When students have prior knowledge, they can interpret the feedback and take corrective actions.

## 2.2.Intelligent tutors

The current and emerging technologies known by artificial intelligence techniques or *intelligent tutors* have an advantage compared to other information technologies which support mathematics instruction [12,18]. Intelligent tutors are developed by combining theories of cognitive science and techniques of artificial intelligence and can be used to provide personalized learning [3,4,6,24]. The *knowledge space theory* is applied to simulate abilities of an expert tutor in tutoring systems, such as ALEKS [14].Built on the foundations of knowledge space theory, the tutor can gauge the level of student's understanding and can detect the correctness of student's next response on the basis of current response. The system can repeat a task sufficient number of times to ensure that student has mastered it then repeat another task, thus generating a interleaved sequence of tasks [34].

## 2.3.Learning Analytics

Learning management systems (LMS), intelligent tutors and other e-Learning systems generate vast amount of user generated data. *Learning analytics* can be applied to provide learner feedback and support, and detect early warning systems [20,22]. The objective of learning analytics is to derive information which can reveal how students use the intelligent tutoring or other e-learning systems and identify potential determinants of academic achievement. [17,20,10].Application of methods of learning analytics can be a powerful means to inform and support learners, teachers and their institutions in better understanding and predicting personal learning needs and performance [20,44]. This analysis can reveal detailed information about different patterns of learning activities, such as the rate at which learner is progressing in the learning environment, under which circumstances their progress is accelerated or decelerated. Cluster analysis, decision trees, artificial neural networks and support vector machines are some examples of techniques used for analysis of such vast amount of data [11]. Cluster analysis involves statistical methods which are commonly used for grouping cases which exhibit similar patterns of learning activities. Cluster analysis techniques group cases and hence can be used as a tool for exploratory data analysis. But this exploratory analysis is done without any subjectivity, hence it provides unambiguous profiles of learning behavior [5,10].

## 3.RESEARCH CONTEXT

ALEKS maintains a record of a set of conceptual units that each student has mastered. This set is termed as the *knowledge state*. A mathematics course may consist of chapters consisting of coherent conceptual units, such as a chapter of *application of percentages* and *geometry*. If *{a}* is a conceptual unit in the list of topics mastered and *{b}* is another conceptual unit not in that list, then the path from *{a} to {b}* isfeasible if the conceptual unit *a* is a pre-requisite of the conceptual unit *b*. The set of all such units like the unit b, form a set of units that the student *is ready to learn* [11,14].

## 3.1.Components of ALEKS

An intelligent tutor system (ITS) has components responsible for each of the following tasks storage and manipulation of domain knowledge, teaching strategies, communication methods and a component which maintains information about student knowledge. It also has the learning component and the control component [23]. The efficiency and sophistication of the first four components determine the relevance of the tutor system in education.

### 3.1.1.Component of domain knowledge

This component is responsible for storage, manipulation and methods of reasoning with the knowledge of some subject domain. ALEKS has developed predefined modules for basic and intermediate courses in arithmetic, algebra, geometry and statistics. ALEKS provides the course content through examples, explanation and problem solving strategies in a textual format. Hyperlinks are embedded within the explanations which can be used for getting additional explanation, such as mathematical definitions. It also uses interactive applets for drawing graphs of equations. There are no issues noted with the accuracy of the content but the following issues are noticed in the presentation of the content.

(a) Ambiguous representation of numeric expressions: The system uses of the symbol '.' to represent multiplication of two numbers, which is read by students as the decimal point. Refer to Figure -1.

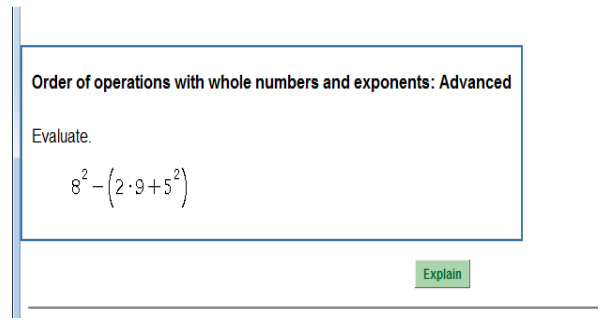

Fig.1:Ambiguous representation of numeric expressions

(b) Static representation of solid figures in geometry: The system uses static representation for three dimensional shapes which is no better than explanation given in any paper based textbook. This is under-utlization of the powerful representation of digital media. It could be enriched with powerful tools like simulation.

### 3.1.2.The teaching component

Any intelligent tutor system is expected to possess an ability to create individualized instructions based on student's background knowledge, level of cognitive development and learning style. [1] as quoted in [23]. ALEKS has the ability to create individualized sequence of topics based on the student's background knowledge and level of cognitive development but the instructions are not delivered using different multimedia format such as audio or video format.

### 3.1.3.Student Knowledge Component

This component is responsible for the diagnostic assessment and modeling of subsequent learning profiles of each student. Upon registering into ALEKS, each student writes an initial or diagnostic assessment. Each student's learning progress is modeled in the form of a Pie. Refer to figure -2. It also maintains logs of all learning activities including topics attempted and topics mastered. Occasional progress tests are administered by ALEKS to detect retention of knowledge. After each progress test, the previous learning score is adjusted. This mechanism provides accurate and up to date model of student's learning progress.

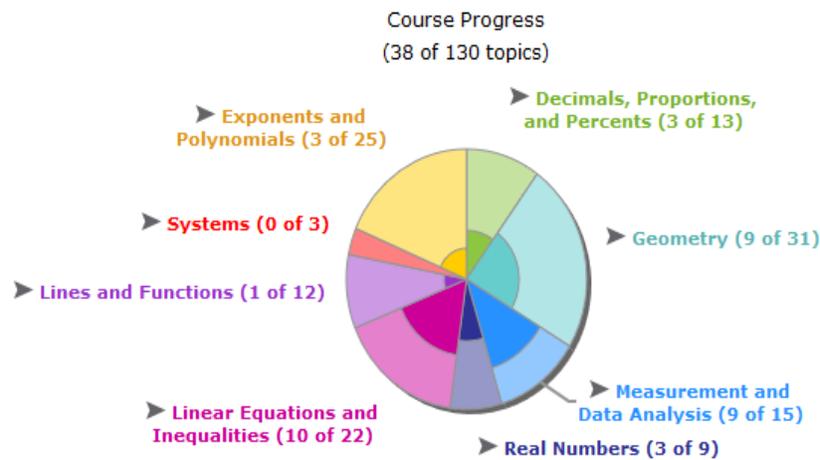

Fig 2:ALEKS pie showing course content and Student progress

### 3.1.4.Communication component

This component is responsible for providing quality interface to students so that student can understand how to use the program and avoid any kind of misinterpretation of any given information. The following issue of misinterpretation is noted. According to the principles knowledge space theory, before attempting any topic a student must master the pre-requisites of that topic. There are not clear instructions presented on the home screen of the system about this fact. If a student has mastered one topic out of 13 from the unit of Measurement, rather than showing remaining 12 topics, only three topics are open for mastering on her Pie. Student often misinterpret this representation as she only has to master three topics. This can be avoided with an explanatory message about the un-attempted prerequisite topics from earlier chapters.

## 4.IMPACT OF ALEKS

In this section, a brief review of strengths and weaknesses of ALEKS in measuring different types of knowledge is presented.

### 4.1.Strengths of ALEKS

The most important feature of ALEKS is that it designs a sequence of activities appropriate for each student and allows the student to learn at his or her own pace. As a result it builds confidence of problem solving [42].

ALEKS has the ability to generate and maintain detailed logs of each student over the complete period of course delivery, which helps the teacher to identify each student's strengths and weaknesses. ALEKS generates and maintains logs of all student activities and shows the up to date status of students' progress. It also shows the three lists for each student describing what the student can do, what is the student ready to learn and the topics attempted but not mastered by the student. It is possible for instructors to monitor students' learning activities periodically from these lists and support students when necessary. Also students can monitor their own learning and progress at their own pace by setting goals.

### 4.2.Limitations of ALEKS

The software can measure student's attainment of factual, semantic and procedural knowledge, but it fails to measure meta-cognition, because student is not expected to demonstrate the strategy used for problem solving. ALEKS assesses student's learning only from the final answer provided to each question. There should be some support for developing metacognition, such as, knowing different strategies to solve a problem and ability to choose the right strategy to solve a problem.

Though frequent progress assessments are administered by the software in order to facilitate learning through assessments, students tend to avoid these automatic progress tests and often request teachers to cancel it for them. It is due to two reasons: the system does not provide detail feedback about the solution submitted during automatic progress tests and they have to relearn all topics which are not retained in the progress test. Some questions in this assessment are taken from the list of topics which a student has not yet mastered but the system finds that the student is ready to learn. Inability to answer these questions, may have a negative effect on their confidence.

### 4.3.Data logs in ALEKS

In this section, a brief description of the ALEKS data logs is presented.

A student can log-in to the system with a purpose of mastering new topics or reviewing topics which she has already mastered. These are recognized by the system as two *modes* of activity, learning and reviewing. The system records details of each student's log-in and log-out times, the time spent by the student on progress assessment, and the mode of each activity. For each student activity, the system keeps a record of the question set by the teaching component and student's response to the question. It also stores the feedback given by the teaching component. The system allows teachers to generate their customized cumulative reports for their classes.

## 5.COURSE STRUCTURE

Two foundation courses covering basic arithmetic, algebra, geometry and statistics are delivered using this software. Students use their iPads to access this program. The software provides explanation and practice problems on each topic. Teachers use the same examples for classroom discussion. Students are expected to master all topics as per their learning pace, which is a formative assessment. There is a possibility of students getting external help in completing the coursework. Therefore a *comprehensive assessment* is administered in class as a summative assessment. This is an individualized progress tests based on what the student has mastered. After this test, the software indicates which topics are retained by the student and which are not retained. In the following discussion the variable *CT* denotes student's marks in this comprehensive assessment and the variable *Pie_Mastery* denotes the total number of topics student mastered before the comprehensive test. Also there are other variables extracted from the data log, which are as follows: student's prior knowledge (denoted by the variable *IA)*, number of topics practiced by the student, total time spent on ALEKS.

## 6.RESEARCH CONTEXT AND METHODOLOGY

In this section, the motivation for this research, the aim of the research and methodology are presented.

A student may log in to the system but do not attempt to master a topic, this *idle time* is included in the total time spent on ALEKS and may not provide accurate information about student's learning. In our preliminary analysis, no correlation was found between the total time spent and the marks in the final exam. A weak positive correlation was found between the variable representing average time per week and the final exam marks (FE). (r=0.177, Sig=0.188, α=0.05)

Refer to the following table 1:

| Correlations | | | |
|---|---|---|---|
| | | Final Exam marksFE | Average time in minutes per week |
| FE | Pearson correlation | 1 | 0.177 |
| | Sig. (2-tailed) | | 0.188 |

Table 1: Output of correlation analysis between *time_spent* and *final exam marks* (denoted by FE)

Also, time taken to master a topic is not significant as students are encouraged to learn at their own pace. The system generated attributes may not provide accurate information about student's learning efforts therefore a derived attribute is defined to investigate learning patterns by taking the ratio of the two variables *number of topics mastered* and *number of topics practiced*. This ratio is represented by the variable *mtop* (which is an abbreviation of *mastered_to_practiced*). It was found that there is a moderately strong, positive and significant correlation between the value of *mtop* and student's final exam marks (r=0.466, sig=0.0) [10]. A high value of this variable indicates that a student is able to master most of the topics she is attempting to master, whereas a low value indicates contrary. It is worth investigating factors which may yield a high value for the variable *mtop* and consequently can facilitate not only learning but also retention of learning.

During initial exploratory investigation, it was found that students with a high value for the variable *mtop* chose their topics sequentially [11]. In the next stage of analysis, data logs of two students were examined. One student had low grades in the CT and another student had high grades in the CT.

ALEKS represents daily activities, such as number of topics practiced and number of topics mastered graphically as shown in figure -3 on the next page.

It can be seen from the above graphs, that student 1 (graph on the left) practiced 22 topics on March 19 but mastered only 11, which indicates a low value for the ratio *mtop*. On the other hand, the student 2 (graph on the right) has mastered almost all topics she practiced. Their learning activities were analyzed in detail, by examining the sequence of topics chosen by each of them. It was revealed that student 1 chose conceptual units from different chapters randomly, whereas student 2 completed units from one chapter and then moved to the next chapter. (Refer to Appendix -1.)

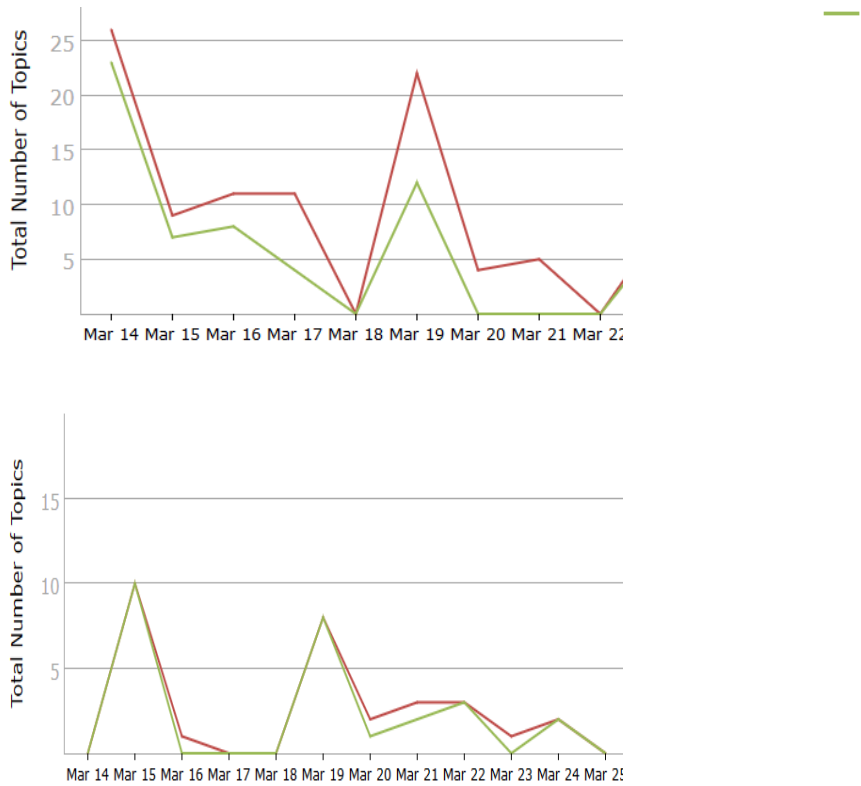

Fig 3: Daily activities of two students ( Number of topics practiced and number of topics mastered.)

This research has the following aims:

1. Develop a model to determine which factors affect academic achievement.
2. Examine students' perceptions about tutoring and cognitive effect of ALEKS and their learning study habits.
3. Examine if their study habits and perceptions have effect on their ability master topics.

In this section description of the research methodology is given.

### 6.1.Data collection

Data was collected from two sources: data logs of ALEKS and student surveys. 58 students participated in the survey. The survey consists of nine questions on the Likert scale from 1 to 5. (1: Strongly disagree; 5: strongly agree).

| IA (score in the initial assessment) | CT (Score in the coursework) | FE (Marks in the final exam) | Time spent in min) | Total number of Topics mastered | Total number of topics practiced | $mtop = \frac{mastered}{practiced}$ |
|---|---|---|---|---|---|---|
| 36% | 83% | 71% | 237 | 22 | 25 | 0.62 |
| 27% | 61% | 61% | 38 | 4 | 4 | 0.80 |
| 28% | 50% | 32% | 48 | 0 | 0 | 0.68 |
| 17% | 65% | 60% | 288 | 26 | 27 | 0.89 |
| 23% | 62% | 60% | 72 | 5 | 5 | 0.90 |

Table 2:Excerpt of the data log

The survey data was triangulated with the data collected from the ALEKS logs, which is described next. A cumulative report was generated from 7-weeks data which included the following information: total time spent, number of topics practiced and number of topics mastered, score in the initial assessment (IA), score in the comprehensive test (CT) and a score in the final exam (FE).

An excerpt of the data log is presented in the table -2. The ratio of these two variables, represented by the variable *mtop* for each week, is used as an indicator of measure of ability to learn independently. The data file consists of the same 58 students who participated in the survey. All students were female students who learn English as their second language.

## 7.RESULTS AND DISCUSSION

This section contains description and results of data analysis methods applied for addressing each research question.

### 7.1 Regression analysis

Multivariate linear regression analysis was done to address research question 1. It was found that though single variable *mtop* is a significant predictor of final exam marks, it only explains 16% of changes in the final exam. [10]. Therefore other predictors were examined through correlation analysis and prior knowledge was found to be another significant predictor.

Students' prior knowledge about the course is represented by the variable IA. The correlation between IA and FE was found to be moderately strong and significant. (r=0.6, Sig=0.0, $\alpha$=0.05). There was no correlation found between IA and *mtop*, therefore further regression analysis was done to examine how effectively these two variables can predict the final exam marks. The results of step-wise regression analysis are shown in the following Table 3.

Table 3:Regression models

|  | Unstandardized Coefficients | t | Sig. |
|---|---|---|---|
| (Constant) | 9.808 | .840 | .405 |
| IA | 0.649 | 4.855 | .000 |
| MeanMTOP | 46.008 | 2.518 | .015 |

Table 4: Regression weights

It can be seen from the table-3 that the model with two predictors shows improved correlation between the predictors (IA and *mtop*) and the dependent variable (FE). ($R^2$ = 42%). The regression weights for independent variables are given in the table 4.

Thus the variable *mtop* is an important predictor of final grades. If the value of this variable is less than 50%, then those students might be at risk of failing. They may need support not just in understanding mathematics concept but also in understanding how to use the system effectively. But it explains only 16% of variation in the final exam marks [10]. Whereas student's prior knowledge alone explains 35% variation in the final exams. Together these two independent variables explain 42% variations in the final exam marks. The non-cognitive factors collected from the surveys, were not found to be significant predictors of the final exam marks.

There are other factors affecting students' ability to learn independently, such as poor language skills and poor technology skills. Poor technology skills result into under-utilization of the features of ALEKS.

### 7.2.Analysis of survey data

Data collected from surveys was tested for reliability and validity to address the research question 2.

Cronbach alpha score of reliability was found to be equal to 0.872, which is a high reliability score. Kaiset-Meyer-Olkin test was applied to check the sampling adequacy and Bartlett's test of sphericity was applied to test inter item correlations. The sample size can be considered adequate for performing factor analysis if the value of the KMO statistics is more than 0.6 and significance value of Batlett's test is less than 0.05 [9,15]. The value of this statistics the significance values were found to be equal to 0.77 and 0.00 respectively. All tests were done using SPSS 23. .

Principal component analysis method was applied with Varimax rotation method. Each item was loaded with a score at least 0.6. Two factors were extracted with 77% cumulative variance explained [15], one factor measured students' perceptions about tutoring effect of ALEKS and the other factor measured students' perceptions about cognitive effect of ALEKS. Refer to the output in Appendix -2. Survey consisted one question about students' study habit. Students chose one of the following two options: I master all topics from one slice then move to the next slice and I pick up any topic from any slice . The first choice indicates that student chooses topics sequentially and the second choice indicates that the student chooses topics randomly. In the following paper these choices are referred to as sequential choice or random choice respectively. From the survey responses, mean score for the two factors were calculated and a binary variable was used to indicate the study habit. (1: Sequential choice, 2: Random choice of topics.)

Cluster analysis is applied to identify groups of people with similar attitudes. All data analysis was done using SPSS 23.

Students were classified into clusters based on the mean values of the two factors identified above: *perception about tutoring effect and perceptions about cognitive effect* and the variable representing study habit. Values more than 3 indicate a positive perceptions about effects of ALEKS. Two-step clustering method was applied since this method is appropriate when the classification variable is continuous and the number of clusters is not known apriori [14]. The software detected three clusters by applying the Log-likelihood method. If total number of topics retained after the comprehensive test (CT) are not significantly less than the total number of topics mastered before the test then, it can be concluded that student is able to retain her learning. Accordingly, research question 3 was formulated into the following null and alternate hypotheses:

$H_{01}$: There is no difference between total number of topics mastered and the CT score for students in cluster 1.
$H_{11}$: Total number of topics mastered and the CT score differed significantly for students in cluster 1.
$H_{02}$: There is no difference between total number of topics mastered and the CT score for students in cluster 2.
$H_{12}$: Total number of topics mastered and the CT score differed significantly for students in cluster 2.
$H_{03}$: There is no difference between total number of topics mastered and the CT score for students in cluster 3.
$H_{13}$: Total number of topics mastered and the CT score differed significantly for students in cluster 3.
Paired sample t-test was applied to test each of these null hypotheses after selecting students from each cluster step-by-step. (Level of significance is set to $\alpha$=0.05 for all tests.)

The table 5 given below shows the cluster profiles, the number of cases in each cluster and results of correlation and paired-sample t-tests for each cluster.

| Cluster number | 1 (Low) | 2 (Medium) | 3 (High) |
|---|---|---|---|
| Method of topic selection | Random | Sequential | Rando |
| Mean score of perceptions about tutoring effect of ALEKS | 2.18 | 4.11 | 4.19 |
| Mean score of perceptions about cognitive effect of ALEKS | 2.65 | 4.00 | 4.07 |
| Number of students in the cluster | 11 | 23 | 24 |

| Mean score of CT | 62 | 71 | 61 |
|---|---|---|---|
| Mean score of number of topics mastered | 71 | 80 | 78 |
| Correlation between CT and number of topics mastered | 0.307 ($\alpha$=0.358) Correlation is not significant | 0.425 ($\alpha$=0.043) Correlation is significant | 0.412 ($\alpha$=0.045) Correlation is significant but slightly weaker than the correlation found in cluster 2. |
| Result of paired sample t-test | t= -0.702 ($\alpha$=0.490) The CT score is less than the total number of topics mastered, but the difference is not significant. | t= -1.6 ($\alpha$=0.125) The CT score is less than the total number of topics mastered, but the difference is not significant. | t= -2.233 ($\alpha$=0.036) The CT score is less than the total number of topics mastered, and the difference is significant. |

Table 5:Cluster profiles

Students in cluster 2 and 3 have positive perceptions about tutoring effects and cognitive effects of ALEKS, but students in cluster 2 chose their topics sequentially whereas students in cluster 3 chose them randomly.

Students in cluster 1 felt that ALEKS did not have any tutoring or cognitive effect on their learning, and they too chose their topics randomly. Overall, this group of students appear to be less motivated and possessed unorganized study habits. They were able to retain their learning but there is no statistical evidence to reject the null hypothesis $H_{01}$. The difference in the number of topics mastered before and after the CT was higher for students in cluster 2 as well, but difference is not statistically significant, therefore the null hypothesis $H_{02}$ is not rejected. The difference in the number of topics mastered before and after the CT was higher for students in cluster 3 is higher than that in other two groups and the difference is statistically significant, therefore the null hypothesis $H_{02}$ is rejected. In summary, it is found that though students who have positive perceptions about tutoring and cognitive effects of ALEKS, if they choose their topics randomly, they may not be able to retain their learning. In case of Students who have overall negative perceptions about tutoring and cognitive effect of ALEKS, there is no statistical evidence to state whether they retained their mastery of learning. But there is evidence to state that avoiding random choice of topics is a useful learning strategy. These findings are consistent with the instructional theory of setting interleaved practice for long term learning in mathematics [34,43]. Instructors can monitor students' learning activities and advise students about sequencing their learning tasks as well as monitor their ability to master all topics that they choose to practice. This guidance will not only improve students' academic achievements, but also will develop a sense of responsibility about their own learning.

## 8.CONCLUSION AND FUTURE DIRECTION

In this paper, we established the ratio of topics mastered to topics practiced *mtop* to measure student's ability to learn independently. This measure and students' prior knowledge are significant predictors of final exam marks. Students' perceptions about tutoring and cognitive effects of ALEKS as well as their methods of topic selection were measured using surveys. Students were classified on the basis of these three non-cognitive factors. This classification formed three groups of students as follows: *negative perceptions and random choice of topics, positive perceptions and random choice of topics and positive perceptions and sequential choice of topics.* A strong positive and significant correlation was found between the course mastery before and after the comprehensive test for students in the cluster three. Though the results are significant, they must be examined on a larger sample in order to generalize. This research will be enhanced by testing the hypotheses on a larger sample.

## REFERENCES

[1] Ackerman, P. L., Sternberg, R. J., & Glaser, R. E. (1989). Learning and individual differences: Advances in theory and research. WH Freeman/Times Books/Henry Holt

[2] ALEKS. (2014). ALEKS, <http://www.aleks.com/> Retrieved on 01.10.2014.

[3] Aleven, V., Roll, I., McLaren, B., & Koedinger, K. (2010). Automated, unobtrusive, action-by-action assessment of self-regulation during learning with an intelligent tutoring system. Educational Psychologist, vol. 45(4), pp. 224-233.

[4] Aleven, V. A., & Koedinger, K. R. (2002). An effective metacognitive strategy: Learning by doing and explaining with a computer-based Cognitive Tutor. Cognitive science, vol. 26(2), pp. 147-179.

[5] Antonenko, P., Toy, S., & Niederhauser, D. (2012). Using cluster analysis for data mining in educational technology research. Educational Technology Research and Development, vol. 60(3), pp. 383-398.

[6] Azevedo, R., & Hadwin, A. (2005). Scaffolding self-regulated learning and metacognition– Implications for the design of computer-based scaffolds. Instructional Science, vol. 33(5), pp. 367-379.

[7] Beal, C., Walles, R., Arroyo, I., & Woolf, B.(2007). On-line tutoring for math achievement testing: A controlled evaluation. Journal of Interactive Online Learning, vol. 6(1), pp. 43-55.

[8] Chen, T. C., Yunus, M., Suraya, A., Ali, W. Z. W., & Bakar, A. (2008). The Effect of an Intelligent Tutoring System (ITS) on Student Achievement in Algebraic Expression. Online Submission, vol. 1(2), pp. 25-38.

[9] Cohen, L., Manion, L. and Morrison, K. (2011) Research Methods in Education (7th Edn). London: Routledge.

[10] Dani (2015). Mining Data-Logs from Intelligent Tutors to Create Learning Profiles of Students, Proceedings of the International Conference on Innovative Engineering and Technologies & Advanced Theoretical Computer Applications, Bangkok, Thailand, Nov 2015.

[11] Dani (2016). Cluster Analysis of data logs generated by intelligent tutor to determine students' learning profiles, 10th International Technology, Education and Development Conference, Valencia, Spain. 7-9 March, 2016.

[12] Dinov, I., Sanchez, J., & Christou, N. (2008). Pedagogical utilization and assessment of the statistic online computational resource in introductory probability and statistics courses. Computers & Education, vol. 50(1), pp.284-300.

[13] Desmarais, M., & Baker, R. (2012). A review of recent advances in learner and skill modeling in intelligent learning environments. User Modeling and User-Adapted Interaction, vol. 22(1-2), pp. 9-38.

[14] Falmagne, J. C., Cosyn, E., Doignon, J. P., & Thiéry, N. (2006). The assessment of knowledge, in theory and in practice. In Formal concept analysis (pp. 61-79). Springer: Berlin Heidelberg.

[15] Field, A. (2009). Discovering Statistics Using SPSS (3rd Edition). London: Sage.

[16] Fyfe, E. R., & Rittle-Johnson, B. (2016). Feedback both helps and hinders learning: The causal role of prior knowledge. Journal of Educational Psychology, vol. 108(1), pp. 82- 98.

[17] Greller, W., & Drachsler, H. (2012). Translating learning into numbers: A generic framework for learning analytics. Journal of Educational Technology & Society, vol 15(3), pp. 42-57.

[18] Hagerty, G., & Smith, S. (2005). Using the Web-based interactive software ALEKS to enhance college algebra. Mathematics & Computer Education, vol. 39(3), pp. 183-194.

[19] Jackson, G. T., Graesser, A. C., & McNamara, D. S. (2009). What Students Expect May Have More Impact Than What They Know or Feel. In AIED pp. 73-80.

[20] Kotsiantis, S., Tselios, N., Filippidi, A., & Komis, V. (2013). Using learning analytics to identify successful learners in a blended learning course. International Journal of Technology Enhanced Learning, Vol. 5(2), pp. 133-150.

[21] Kulik, C. L. C., Kulik, J. A., & Bangert-Drowns, R. L. (1990). Effectiveness of mastery learning programs: A meta-analysis. Review of educational research, vol. 60(2), pp. 265-299.

[22] Libbrecht, P., Rebholz, S., Herding, D., Müller, W., & Tscheulin, F. (2012). Understanding the Learners' Actions when Using Mathematics Learning Tools. In Intelligent Computer Mathematics (pp. 111-126). Springer Berlin Heidelberg.

[23] Mark, M., & Greer, J. (1993). Evaluation methodologies for intelligent tutoring systems. Journal of Artificial Intelligence in Education, vol. 4, pp. 129-129.

[24] McArthur, D., & Stasz, C. (1990). An intelligent tutor for basic algebra. R-3811- NSF, RAND Corporation, Santa Monica, CA.

[25] McDaniel, M. A., Fadler, C. L., & Pashler, H. (2013). Effects of Spaced Versus MassedTraining in Function Learning. Journal of Experimental Psychology: Learning, Memory, andCognition. Advance online publication. doi: 10.1037/a0032184

[26] McGatha, M., & Bush, W. (2013). Classroom assessment in mathematics. In J. McMillan (Ed.), SAGE handbook of research on classroom assessment. (pp. 448-461). Thousand Oaks, SAGE Publications, Inc.: CA

[27] Melis, E., & Siekmann, J. (2004). Activemath: An intelligent tutoring system for mathematics. In Artificial Intelligence and Soft Computing-ICAISC 2004 (pp. 91-101). Springer: Berlin Heidelberg.

[28] Miller, T. (2009). Formative computer  based assessment in higher education: The effectiveness of feedback in supporting student learning. Assessment & Evaluation in Higher Education, vol. 34(2), pp. 181-192.

[29] Nguyen, D, Hsieh, Y., & Allen, G. (2006). The impact of web-based assessment and practice on students' mathematics learning attitudes. Journal of Computers in Mathematics and Science Teaching, vol. 25(3), pp. 251-279.

[30] Nguyen, D. M., Hsieh, Y. C., & Allen, G. D. (2006). The impact of web-based assessment and practice on students' mathematics learning attitudes. Journal of Computers in Mathematics and Science Teaching, 25(3), pp.251-279.

[31] Nicol, D. (2006). Increasing success in first year courses: Assessment re-design, self-regulation and learning technologies. In Proceedings of the 23rd annual ascilite conference.

[32] Palocsay, S., & Stevens, S. (2008). A Study of the Effectiveness of Web  Based Homework in Teaching Undergraduate Business Statistics. Decision Sciences Journal of Innovative Education, vol. 6(2), pp. 213-232.

[33] Rau, M. A., Aleven, V., & Rummel, N. (2010). Blocked versus interleaved practice with multiple representations in an intelligent tutoring system for fractions. In Intelligent tutoring systems (pp. 413-422). Springer Berlin Heidelberg.

[34] Rohrer, D., Dedrick, R. F., & Stershic, S. (2015). Interleaved practice improves mathematics learning. Journal of Educational Psychology, vol. 107(3), pp. 900-908.

[35] Rohrer, D., Dedrick, R. F., & Burgess, K. (2014). The benefit of interleaved mathematics practice is not limited to superficially similar kinds of problems. Psychonomic bulletin & review, 21(5), pp.1323-1330.

[36] Sana, F., Yan, V. X., & Kim, J. A. (2016). Study Sequence Matters for the Inductive Learning of Cognitive Concepts, Journal of Educational Psychology. Advance online publication.http://dx.doi.org/10.1037/edu0000119.

[37] Shute V., (2008). Focus on Formative Feedback. Review of Educational Research, vol. 78(1), pp. 153-189.

[38] Shute, V., & Underwood, J. (2006). Diagnostic assessment in mathematics problem solving. Technology instruction cognition and learning, vol. 3(1/2), pp. 151-166.

[39] Siemens, G., & Long, P. (2011). Penetrating the Fog: Analytics in Learning and Education. EDUCAUSE review, vol 46(5), pp. 30-36.

[40] Stiggins, R., & Chappuis, J. (2005). Using student-involved classroom assessment to close achievement gaps. Theory into practice, vol. 44(1), pp.11-18.

[41] Tempelaar, D. T., Rienties, B., & Giesbers, B. (2015). In search for the most informative data for feedback generation: Learning Analytics in a data-rich context. Computers in Human Behavior, 47, pp. 157-167.

[42] Wood, D., & Wood, H. (1996). Vygotsky, tutoring and learning. Oxford review of Education, vol. 22(1), pp. 5-16.

[43] Xiong, X., & Beck, J. E. (2014). A study of exploring different schedules of spacing and retrieval interval on mathematics skills in ITS environment. In Intelligent tutoring systems (pp. 504-509). Springer International Publishing.

[44] Yu, T., & Jo, I. H. (2014). Educational technology approach toward learning analytics: Relationship between student online behavior and learning performance in higher education. In Proceedings of the Fourth International Conference on Learning Analytics and Knowledge (pp. 269-270). ACM.

**APPENDIX -1**

*Illustration of pre-requisite relationship between conceptual units and possible learning paths*

a,b,c,d,e,f,g,h denote conceptual units. They are organized into chapters as shown below. Arrow between two units indicates the pre-requisite relationship between those conceptual units.

a is a pre-requisite for b, c and d.

d is a pre-requisite for e

f and g have no pre-requisites

g is a pre-requisite for h

*Learning path 1*: a⟶ b⟶ d⟶ f ;  *Learning path 2:* f ⟶ g ⟶a ⟶d

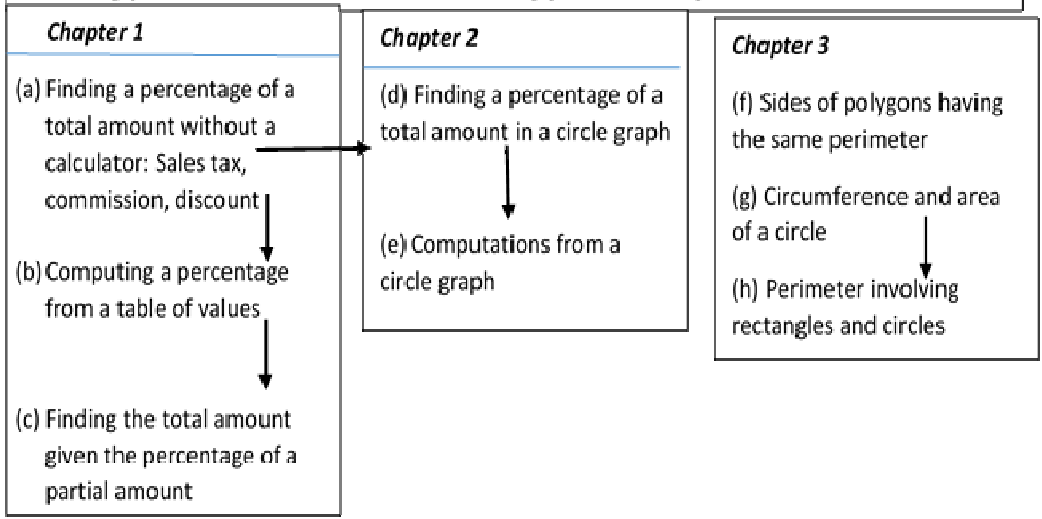

**Chapter 1**

(a) Finding a percentage of a total amount without a calculator: Sales tax, commission, discount

(b) Computing a percentage from a table of values

(c) Finding the total amount given the percentage of a partial amount

**Chapter 2**

(d) Finding a percentage of a total amount in a circle graph

(e) Computations from a circle graph

**Chapter 3**

(f) Sides of polygons having the same perimeter

(g) Circumference and area of a circle

(h) Perimeter involving rectangles and circles

Example of sequential choice of attempt to master conceptual units :

a ⟶ b ⟶ c ⟶ d ⟶ e ⟶f

Attempt to master all units of chapter 1 and then attempt to master all units of chapter 2.

Problem solving strategies for units *a,b,c,d,e* are not different.

Example of random selection of conceptual units:

a⟶d⟶f ⟶ g⟶b⟶c ⟶e

Attempt to master units from chapter 2 before mastering all units from chapter 1.

Units *a* and *d* require different problem solving strategies. Similarly units *d* and *f* require different problem solving strategies.

**APPENDIX -2**

SPSS output of factor analysis

### Rotated Component Matrix[a]

|  | Component | |
|---|---|---|
|  | 1 | 2 |
| ALEKS encouraged me to think for myself. | .909 | |
| ALEKS encouraged the development of my knowledge. | .853 | |
| ALEKS made helpful comments on my work. | .867 | |
| ALEKS provided helpful feedback on my work. | .798 | |
| ALEKS sensed when I needed help. | | .539 |
| ALEKS helped me master topics in advance. | | .649 |
| ALEKS provided me with detailed explanation. | | .840 |
| ALEKS is easy to use. | | .905 |
| ALEKS is enjoyable and stimulating | | .661 |

Extraction Method: Principal Component Analysis.

 Rotation Method: Varimax with Kaiser Normalization.

a. Rotation converged in 3 iterations.